\input harvmac

\overfullrule=0pt
\def\Title#1#2{\rightline{#1}\ifx\answ\bigans\nopagenumbers\pageno0\vskip1in
\else\pageno1\vskip.8in\fi \centerline{\titlefont #2}\vskip .5in}

\font\ticp=cmcsc10

\font\secfont=cmcsc10

%
%defs
%
\baselineskip=18pt plus 2pt minus 2pt

\def\ajou#1&#2(#3){\ \sl#1\bf#2\rm(19#3)}
%
%Caligraphic
\def\CH{{\cal H}}

%
%\def\eightbig#1{{\hbox{$\textfont0=\ninerm\textfont2=\niesy\left#1\vbox
% to6.5pt{}
%\right.\n@space$}}}
%\eightbig{N}

%Roman

%

%Greek
\def\x{\xi}

\def\s{\sigma}
\def\g{\gamma}
\def\t{\tau}
\def\a{\alpha}
\def\b{\beta}
\def\d{\delta}

\def\p{\pi}

\def\l{{\lambda}}
\def\O{{\Omega}}

%

%This paper

\def\TrH#1{ {\raise -.5em
                      \hbox{$\buildrel {\textstyle  {\rm Tr } }\over
{\scriptscriptstyle \CH _ {#1}}$}~}}

\def\IZ{\relax\ifmmode\mathchoice
{\hbox{\cmss Z\kern-.4em Z}}{\hbox{\cmss Z\kern-.4em Z}}
{\lower.9pt\hbox{\cmsss Z\kern-.4em Z}}
{\lower1.2pt\hbox{\cmsss Z\kern-.4em Z}}\else{\cmss Z\kern-.4em Z}\fi}
\def\IC{\relax\hbox{$\inbar\kern-.3em{\rm C}$}}
\def\IR{\relax{\rm I\kern-.18em R}}
\def\1{\relax 1 { \rm \kern-.35em I}}
\font\cmss=cmss10 \font\cmsss=cmss10 at 7pt

\def\OF{{\Omega (-1)^{F_L} R_2}}
%

%Shortforms
\def\frac#1#2{{#1 \over #2}}

\def\p+{{\partial_+}}

\def\half{{1 \over 2}}

%

%refs

%

\Title{\vbox{\baselineskip12pt
\hbox{\ticp caltech preprint}
\hbox{\ticp doe research and}
\hbox{\ticp development report}
\hbox{}
\hbox{hep-th/9607041}
}}
{\vbox{\centerline {\bf A NOTE ON ORIENTIFOLDS AND F-THEORY} }}

\centerline{{\ticp
Atish Dabholkar\footnote{$^1$}{e-mail: atish@theory.caltech.edu}
and Jaemo Park\footnote{$^2$}{e-mail: jpk@theory.caltech.edu}
}}

\vskip.1in
\centerline{\it Lauritsen Laboratory of  High Energy Physics}
\centerline{\it California Institute of Technology}
\centerline{\it Pasadena, CA 91125, USA}

\vskip .1in

\bigskip
\centerline{ABSTRACT}
\medskip

An orientifold of Type-IIB theory on a $K3$ realized as a
$Z_2$ orbifold is constructed which corresponds to F-theory
compactification on a Calabi-Yau orbifold with Hodge numbers
$(51, 3)$. The T-dual of this model is analogous
to an orbifold with discrete torsion in that the action of
orientation reversal has an additional phase on the twisted
sectors, and both 9-branes and 5-branes carry orthogonal
gauge groups. An orientifold of the $Z_3$ orbifold and its
relation to F-theory is briefly discussed.

\bigskip

\bigskip
%\baselineskip=20pt plus 2pt minus 2pt
%\draft
\Date{July, 1996}

\vfill\eject

%References
\def\npb#1#2#3{{\sl Nucl. Phys.} {\bf B#1} (#2) #3}
\def\plb#1#2#3{{\sl Phys. Lett.} {\bf B#1} (#2) #3}
\def\prl#1#2#3{{\sl Phys. Rev. Lett. }{\bf #1} (#2) #3}

\def\cmp#1#2#3{{\sl Comm. Math. Phys. }{\bf #1} (#2) #3}
\def\mpl#1#2#3{{\sl Mod. Phys. Lett. }{\bf #1} (#2) #3}

%-------------------
% references
%-------------------
%
\lref\Orie{
A. Sagnotti, in Cargese '87, ``Non-perturbative Quantum
Field Theory,'' ed. G. Mack et. al. (Pergamon Press, 1988) p. 521\semi
P. Horava, \npb{327}{1989}{461}\semi
J. Dai, R. G. Leigh, and J. Polchinski,
\mpl{A4}{1989}{2073}.}

\lref\Sagn{M. Bianchi and A. Sagnotti,
\plb{247}{1990}{517}; \npb{361}{1991}{519}}

\lref\GoMu{R.~Gopakumar and S.~Mukhi, private communication.}

\lref\DMW{M.~J.~Duff, R.~Minasian, and E.~Witten,
``Evidence for Heterotic/Heterotic Duality,''
CTP-TAMU-54/95, hep-th/9601036.}

\lref\DaPaI{
A. Dabholkar and J. Park, ``An Orientifold of Type IIB theory on K3,''
CALT-68-2038, hep-th/9602030.}

\lref\DaPaII{
A. Dabholkar and J. Park, ``Strings on Orientifolds,'' CALT-68-2051,
hep-th/9604178.}

\lref\GiJoI{
E. Gimon and C. Johnson, ``K3 Orientifolds,'' NSF-ITP-96-16, hep-th/9604129.}

\lref\GiJoII{
E. Gimon and C. Johnson, ``Multiple Realizations of N=1 Vacua in
Six-Dimensions,'' NSF-ITP-96-55, hep-th/9606176.}

\lref\SenI{A. Sen, ``M-Theory on $(K3 \times S^1 )/Z_2$,'' MRI-PHY/07/96,
hep-th/9602010.}

\lref\SenII{A. Sen, ``Orbifolds of M-Theory and String Theory,''
MRI-PHY/10/96, hep-th/9603113.}

\lref\SenIII{A. Sen, ``F Theory and Orientifolds,'' MRI-PHY/14/96,
hep-th/9605150.}

\lref\Vois{C. Voisin, Journ\'{e}es de G\'{e}om\'{e}trie Alg\'{e}brique
d'Orsay(Orsay,1992),
Ast\'{e}risque No. 218(1993), 273.}

\lref\Borc{C. Borcea, ``K3 Surfaces with Involution and Mirror Pairs of
Calabi-Yau Manifolds,'' in {\it Essays on Mirror manifolds} {\rm Vol. II},
to appear.}

\lref\Aspi{P. Aspinwall, \npb{460}{1996}{57}.}

\lref\VafaI{C. Vafa, ``Evidence for F Theory,''
HUTP-96-A004, hep-th/9602022.}

\lref\VafaII{C.~Vafa, ``Modular Invariance and
Discrete Torsion on Orbifolds,'' \npb{273}{1986}{592}.}

\lref\DHVW{L.~Dixon, J.~Harvey, C.~Vafa, and E.~Witten,
``Strings on Orbifolds I and II,'' \npb{261}{1985}{678};
\npb{274}{1986}{285}.}

\lref\Zasl{E.~Zaslow, ``Topological Orbifold Couplings
and Quantum Cohomology Rings,'' \cmp{156}{1993}{301}.}

\lref\VaWi{C. Vafa and E. Witten,
{\sl Jour. Geom. Phys.}{\bf 15} (1995) 189.}

\lref\MoVaI{D. Morrison and C. Vafa,
``Compactifications of F-Theory on
Calabi-Yau Threefolds-I,'' hep-th/9602114.}

\lref\MoVaII{D. Morrison and C. Vafa,
``Compactifications of F-Theory on
Calabi-Yau Threefolds-II,'' hep-th/9603161.}

\lref\PolcI{J. Polchinski, \prl{75}{1995}{4724}.}

\lref\PolcII{J. Polchinski, ``Tensors From K3 Orientifolds,''
NSF-ITP-96-54, hep-th/9606165.}

\lref\PCJ{J. Polchinski, S. Chaudhuri and C. Johnson,
``Notes on D-Branes,''
 NSF-ITP-96-003, hep-th/9602052.}

\lref\GiPo{E.~G.~Gimon and J.~Polchinski, ``Consistency Conditions
for Orientifolds and D-manifolds,'' hep-th/9601038.}

\lref\Witt{E.~Witten, ``Five-branes and M-Theory on an Orbifold,''
hep-th/9512219.}

\lref\SeWi{N.~Seiberg and E.~Witten, ``Comments on String
Dynamics in Six Dimensions,'' RU-96-12, hep-th/9603003.}

\lref\WittI{E.~Witten, ``Phase Transitions in M-theory and F-theory,
IASSNS-HEP-96-26, hep-th/9603150.}

\lref\WittII{E.~Witten, ``Small Instantons in String Theory,''
\npb{460}{1996}{541}.}

\lref\Six{M.~Berkooz, R.~G.~Leigh, J.~Polchinski,
J.~H.~Schwarz, N.~Seiberg, and E.~Witten,
``Anomalies, Dualities, and Topology of $D=6$ $N=1$
Superstring Vacua,'' hep-th/9605184.}

\lref\DMW{M.~J.~Duff, R.~Minasian, and E.~Witten,
``Evidence for Heterotic/Heterotic Duality,''
CTP-TAMU-54/95, hep-th/9601036.}

Orientifolds are a generalization of orbifolds in which
the orbifold symmetry is a combination of a spacetime symmetry
and orientation reversal on the worldsheet
\refs{\Orie, \PolcI, \GiPo, \PCJ}. These techniques
have significantly enlarged the set of string vacua that can be
studied perturbatively. Several new string vacua
can now be constructed as orientifolds
which exhibit novel dynamical phenomena and have interesting
nonperturbative duals in
M-theory, F-theory, or heterotic string theory.

One important application of orientifolds is in the construction
of models in six dimensions with $N=1$ supersymmetry.
The dynamics of these theories offers many surprises like
the appearance of tensionless strings which can cause a phase transition
in which the number of tensor multiplets changes \refs{\DMW, \SeWi, \WittI},
or the appearance of enhanced gauge symmetry when an instanton
shrinks to zero scale size \WittII.
Orientifolds are useful in understanding some aspects of these
phenomena perturbatively. For instance, the models with multiple
tensor multiplets are inaccessible using usual Calabi-Yau
compactifications which give only a single tensor multiplet.
However, one can easily construct
orientifolds \refs{\Sagn, \DaPaI, \GiJoI, \DaPaII, \GiJoII}
with multiple tensor multiplets at special points in this
moduli space. By turning on the moduli in the tensor multiplets
one can move away from these special points and thus explore different
regions of the moduli space that are separated by phase boundaries.
Some of these models \DaPaI\ are known to have
M-theory duals\refs{\SenI, \SenII}.
The extra tensor multiplets which arise in M-theory from the addition of
M-theory 5-branes occur perturbatively in the dual orientifold.
Similarly, small instantons, which cannot be described
as a conformal field theory in heterotic compactifications,
have a perturbative description in terms of a Dirichlet 5-brane
in the dual orientifold \refs{\WittII, \Six}.
In particular, the enhanced $Sp(k)$ symmetry when $k$ small
instantons coincide can be understood in terms of coincident
5-branes with a specific symplectic projection in the open string
sector that is determined by the consistency of the world-sheet
theory.

Another more recent application of orientifolds is in connection
with F-theory \refs{\VafaI, \MoVaI, \MoVaII}.
F-theory refers to a new way of compactifying
Type-IIB theory in which the complex coupling $\l$ of
Type-IIB theory is allowed to vary over space. The coupling
is given by $\l = \x + i e^{-\phi}$ where $\phi$ is the dilaton
from the NSNS sector and $\x$ is the RR scalar. Consider
an elliptically fibered Calabi-Yau manifold $K$ which is a fiber
bundle over a base manifold $B$ with a torus as a fiber whose
complex structure parameter is $\t$. Even-though $K$ is a smooth
manifold, there will be points in the base manifolds where
the fiber becomes singular, and the parameter $\t$ can have
a nontrivial $SL(2, Z)$ monodromy around these points. An
F-theory compactification on $K$ refers to a compactification
of Type-IIB theory on $B$, where the coupling $\l$ is
identified with $\t$. The nontrivial monodromy of $\l$ around
the singular points then means that there are 7-branes at those points
that are magnetically charged with respect to the scalar $\l$. Typically,
the base manifold is not Ricci-flat and moreover, because $\l$
is varying, there is a nonvanishing RR background. These backgrounds
cannot, therefore, be described using conformal field theory.
For special choices of the manifolds $K$, however,
an F-theory compactification is equivalent
to a perturbative Type-IIB orientifold.
This follows from an observation due to Sen \SenIII\ that
the element $-\1$
of $SL(2, Z)$ which is not an element of $PSL(2, Z)$
is a perturbative symmetry of Type-IIB. It flips the sign
of the two 2-form fields $B^1_{MN}$ and $B^2_{MN}$,
but leaves all other massless fields, in particular, the
coupling fields $\l$ invariant.
{}From its action on the massless fields it is easy to check
that this element represents the action of
$\Omega (-1)^{F_L}$ where $\Omega$
is orientation reversal on the worldsheet and $F_L$ is the
spacetime fermion number of the left-movers.
In the example considered by Sen, $K$ is a $K3$ surface
that is a $Z_2$ orbifold of a four-tours; F-theory on this
surface corresponds to a Type-IIB orientifold with the
orientifold group $\{ 1, \Omega (-1)^{F_L} \sigma\}$ where
$\sigma$  is a specific $Z_2$ involution of $K3$, and
is T-dual to Type-I theory.
Such an identification of F-theory with an orientifold is very
useful. For instance, it was used in \SenIII to establish the
duality between F-theory on $K3$ and the heterotic string on $T^2$
by relating it to the duality between the Type-I and the heterotic
string in ten dimensions.

In this note we analyze an orientifold of a $K3$ orbifold
which gives $N=1$ supersymmetry in six dimensions. Its
T-dual has the same orientifold group
as the Type-I orientifold analyzed by Gimon and Polchinski \GiPo,
but the orientation reversal symmetry $\Omega$ acts
with an additional minus sign on the twisted sector states of the
orbifold. One is familiar with an analogous situation in orbifold
constructions.
For a $Z_k \times Z_k$ orbifold symmetry,
there are $k$ inequivalent orbifolds which correspond to
turning on discrete torsion\refs{\VafaII, \VaWi}.
These different orbifolds correspond
to the $k$ distinct choices of phases for the action
of the generator of one $Z_k$ subgroup on the sectors twisted by
other generators.

This model illustrates interesting new features that
are relevant to all the applications mentioned earlier:
the unusual action of orientation reversal gives rise to multiple tensor
multiplets,
the 5-branes at the fixed points of the orbifold
have orthogonal projection instead of the symplectic projection of a
small instanton at a nonsingular point, and it is perturbatively
equivalent to F-theory on a Calabi-Yau orbifold $T^6/\{ Z_2 \times Z_2\}$
with Hodge numbers $(h^{11}, h^{21})= (51, 3)$ \VaWi.
Using the formulae in \MoVaI\ we see that this
F-theory compactification gives $17$ tensor multiplets,
four neutral hypermultiplets,
$SO(8)^8$ gauge group, and no charged hypermultiplets.
Our aim in the following is to see how the orientifold
reproduces this spectrum.

Let us denote the complex coordinates of the six-torus
by $z_1, z_2, z_3$ with
identifications $z_l \equiv z_l + 1 \equiv z_l + i,
l=1, 2, 3$. The $Z_2 \times Z_2$ symmetry is generated by the
elements $\a$ and $\b$ where
\eqn\orbi{\eqalign{
\a \quad : & \quad (z_1, z_2, z_3) \rightarrow (-z_1, -z_2, z_3),\cr
\b \quad : & \quad (z_1, z_2, z_3) \rightarrow (z_1, -z_2, -z_3).\cr}}
It is easy to work out the cohomology \refs{\DHVW, \Zasl, \VaWi}.
The untwisted sector
contributes $(3, 3)$ to $(h^{11}, h^{21})$, and the sectors
twisted by $\a$, $\b$, and $\a\b$ each contribute $(16, 0)$, giving
$(51, 3)$ altogether. To obtain the corresponding orientifold, we
take $z_3$ as the coordinate of the fiber, and  consider Type-IIB
compactified on a four-torus with coordinates
$(z_1, z_2)$: $z_1 = X^6 +iX^7$, $z_2 = X^8 +iX^9$.
Orbifolding with the symmetry $\a$ gives Type-IIB on $K3= T^4/Z_2$.
The element $\b$ can be written as $R_2 R_3$ where
$R_2$ is a geometric symmetry $(z_1, z_2) \rightarrow (z_1, -z_2)$,
and $R_3$, which reflects the fiber, is nothing but the element
$-\1$ of $SL(2, Z)$ which
corresponds to the operation $\Omega (-1)^{F_L}$
as explained in the preceding paragraph.
We are thus led to consider an orientifold of Type-IIB on
$K3$ with the orientifold group $\{1, \Omega (-1)^{F_L} R_2\}$
\foot{We would like to thank S.~Mukhi for this observation which
prompted this investigation \GoMu.}.

This orbifold is a special case of a large
class of elliptic Calabi-Yau threefolds
studied by Voisin \Vois\ and Borcea \Borc\ and discussed
in \refs{\Aspi, \MoVaII}.
One can take the base to be a $K3$ which admits an
involution $\s$ under which the holomorphic 2-form $\omega$ is odd,
and construct the Calabi-Yau as an orbifold
$K3 \times T^2/\{1, \s R_3\}$ where $R_3$ is the
reflection of the torus.
It should be possible to generalize the considerations of this
paper to this whole class of models.

The projection that we wish to perform is
$\frac{1}{4}(1 + \Omega (-1)^{F_L} R_2) ( 1 + R)$ where $R= R_1 R_2$.
The projection $\half (1 + R)$ gives us Type-IIB theory on
a $K3$ which has $21$ tensor multiplets of $N=2$
supersymmetry which is sum of a tensor multiplet and
a hypermultiplet of $N=1$ supersymmetry.  Five of these multiplets
come from the untwisted sector, and the remaining $16$ come from
the twisted sectors at the $16$ fixed points of the orbifold.
Now, from the arguments of \refs{\DaPaII, \SenIII}, one would
have expected, by T-duality in the $89$ directions, that the operation
$\OF$ is equivalent to the operation $\Omega$. It seems, therefore,
that we get an orientifold of $T^4$ with the orientifold group
$\{ 1, R, \Omega, \Omega R\}$ which is nothing but a Type-I
orientifold on $K3$ analyzed by \GiPo. The massless spectrum,
however, is very different; for example, the closed-string
spectrum of the model of \GiPo\
has only one tensor multiplet instead of $17$, and
$20$ neutral hypermultiplets instead of four.
The reason for this mismatch is that,
even though the two projections
are the same in the untwisted sector, they
are different in the twisted sectors of the orbifold.
This is clear if we look at the action of $\Omega (-1)^{F_L} R_2$ on
the twisted sectors.
The operation $\Omega$ that is dual to
$\Omega (-1)^{F_L} R_2$ corresponds to $\Omega_0 T$, where
$\Omega_0$ is the operation considered in \GiPo, and $T$
is a symmetry of the orbifold that flips the sign of the
twist fields at all fixed points.
In untwisted sector in both theories give
one tensor multiplet and four hypermultiplets.
But in the twisted sector at each fixed point,
$\Omega_0$ projects out the tensor multiplet
and keeps the hypermultiplet
giving the closed string spectrum of Type-I on $K3$
whereas $\Omega$
keeps the tensor multiplet and projects out
the hypermultiplet giving $17$ tensor multiplets and four
hypermultiplets altogether, as required.

Let us now turn to the open-string sector.
We shall follow the notation of \GiPo\ in the T-dual picture
so that we have 7-branes and 7'-branes instead of
9-branes and 5-branes respectively.
The T-dual picture turns out to be easier because then the
symmetry
breaking is given by geometric separation between
branes instead of by Wilson lines. The orientifold group
in this case is $\{ 1, R, \Omega(-1)^{F_R}R_1, \OF\}$.
Note that both $R_1$ and $R_2$,
and similarly $\Omega (-1)^{F_L}$ and $\Omega (-1)^{F_R}$
all square to $(-1)^F$ but the elements of the orientifold
group all square to $\1$ as they should. To simplify the
notation, let us denote $\Omega(-1)^{F_R}R_1$ and
$\OF$ by $\Omega_1$ and $\Omega_2$ respectively.
To determine the open-string sector we need to determine,
as in \GiPo,
the number of branes of each type and the eight $\gamma$
matrices that give the action of the four orientifold group
elements on $7$ and $7'$ branes.

Before discussing the details of the calculation let us present
the results. Tadpole cancellation requires $32$ branes
of each kind; the $32$ 7-branes are located
at the four fixed planes of $R_1$ in groups of eight,
and the $32$  7'-branes are located at the fixed planes
of $R_2$ in groups of eight. Moreover, by a unitary
change of basis, the various gamma matrices are given by
\eqn\solution{\eqalign{
&\g_{1, 7} = \1, \quad \g_{\O_1, 7} = \1, \quad \g_{R, 7} = \1,
\quad \g_{\O_2, 7} = \1;\cr
&\g_{1, 7'} = \1, \quad \g_{\O_2, 7'} = \1, \quad \g_{R, 7'} = -\1,
\quad \g_{\O_1, 7'} = -\1.\cr
}}

Now consider the massless bosonic states coming from the
$77$ sector at the fixed point where eight 7-branes are located.
The vectors are given by
$\psi_{-1/2}^\mu|0,ij\rangle \lambda_{ji}, \mu = 1, 2, 3, 4$;
$R=+1$ implies $\l =\l$, and $\O_1 =+1$ implies $\l = -\l^T$,
which means that the vectors are in the adjoint of $SO(8)$.
The scalars are given by
$\psi_{-1/2}^\mu|0,ij\rangle \lambda_{ji}, \mu = 6, 7, 8, 9$;
$R=+1$ implies $\l = -\l$, which means that they are
all projected out. {}From four fixed planes
of $R_1$ we get $SO(8)^4$, and similarly from the
$7'7'$ sector we get another $SO(8)^4$.
Thus, altogether we get $SO(8)^8$ with no charged
hypermultiplets.

In the $77'$ sector there is a subtlety. In this case,
we have to choose the oscillator vacuum of this sector
to be odd under the action of $R$ instead of even as in
\GiPo. This is consistent with factorization because
$77'$ and $7'7$ states can turn into $77$ or $7'7'$ states,
but we cannot have two $77$ or two $7'7'$ states turning
into a $77'$ state. So one can choose
the $77'$ vacuum to be odd and the $77$ and $7'7'$
vacua to be even.
We shall explain two paragraphs later
that this choice is indeed forced upon us by consistency.
In this sector the fermions $\Psi^m$
have integer modings, so the
ground states are given by a representation of Clifford
algebra generated by the zero modes. The total state
after GSO projection is $|s_3, s_4, ij\rangle \l_{ji},
s_3 = -s_4$ where $s_3, s_4 = \pm\half$.
We choose $R$ on these GSO-projected vacuum states to be $-1$
instead of $+1$. Thus, R=+1 on the total state
implies $\l = -\l$ which projects out
the massless states completely.
To summarize, we get $17$ tensor multiplets and four
hypermultiplets from the
closed-string sector, and $SO(8)^8$ gauge group with no charged
hypermultiplets from the open-string sector, altogether
in agreement with the F-theory spectrum.

This determination of the spectrum, however, poses the following
puzzle. {}From arguments similar to those presented in \GiPo,
one would have expected that if $\g_{\O_1, 7}$ is symmetric then
$\g_{\O_1, 7'}$ should be antisymmetric. How did we then
obtain a solution in which both are symmetric? To see
that this is a consistent choice, let us recall the argument of \GiPo.
In the following we shall often switch between our model
and its T-dual.
In order to obtain a true representation (and not merely a projective
representation) of the orientifold symmetry that
we are gauging, we must have $\O^2=\1$ in
the full string Hilbert space, which is a direct product
of the Fock space of string oscillators and the Chan-Paton index space.
Now, because $\Omega^2$ is $-1$ on the oscillator
part of the massless states, it must be compensated by choosing
$-1$ on the Chan-Paton part.  This forces
$\g_{\O , 5}$ to be antisymmetric if $\g_{\O , 9}$ is
symmetric. In our case, however, because of our choice of
$R = -\1$ on the GSO-projected vacuum states
that we used in the previous
paragraph, the massless states in the ${59}$
are projected out. Moreover, it is easy to see that at the massive
level, the oscillator part of the physical states that are left after
the GSO and the R-projection all have $\O^2=+1$.
This is so because only the states at half-integer mass levels
survive the projections. Now, $\Omega^2 = -1$ for the half
integer oscillator modes, and moreover because
$\Omega^2 =-1$ on the oscillator vacuum as noted
by \GiPo, the total oscillator state has $\O^2 =+1$.
This in turn implies that in the Chan-Paton space we
must choose $\Omega^2 =1$ which means
that if $\g_{\O, 9}$ is symmetric then $\g_{\O, 5}$
must also be symmetric.
To put it differently, of the whole tower of states in
the $59$ sector, the states that are kept after the
GSO and the R projection, all have $\O^2 =-1$ in \GiPo,
but have $\O^2 =+1$ in this paper. Thus, the choice of
the projection $R$ and the sign of the eigenvalue of
$\O^2$ are correlated. Under T-duality $59$ sector
corresponds to $7'7$ sector and the argument above can
be repeated there.

Let us now show that the spectrum described above satisfies
all consistency requirements, and is moreover uniquely determined.
Tadpole calculation in this case is very similar to the T-dual of \GiPo.
The Klein-Bottle and the M\"obius strip amplitudes are identical,
and for the cylinder amplitude, the only difference is the
additional minus sign in the $77'$ and $7'7$ sector in calculating
the trace of $R$. The tadpoles are thus given,
in the notation of \GiPo, by
\eqn\tadpole{\eqalign{
&\frac{v_6 v_2}{16 v'_2}
\left\{ 32^2 - 64 {\rm Tr}(\gamma_{\Omega_1, 7}^{-1}\gamma_{\Omega_1, 7}^T)
+ ( {\rm Tr}(\gamma_{1, 7}) )^2 \right\}\cr
+&\frac{v_6 v'_2}{16 v_2}
\left\{ 32^2 - 64 {\rm Tr}(\gamma_{\Omega_2, 7'}^{-1}\gamma_{\Omega_2, 7'}^T)
+ ({\rm Tr}(\gamma_{1, 7'}))^2 \right\}\cr
+&\frac{v_6}{8}\left\{ {\rm Tr}(\gamma_{R,7}){\rm Tr}(\gamma_{R,7'})
+2 \sum_{I=1}^{4} \left({\rm Tr}(\gamma_{R,7})\right)^2
+2 \sum_{I'=1}^{4} \left({\rm Tr}(\gamma_{R,7'}) \right)^2 \right\}.\cr
}}
Here $v_6$ is the regularized volume of the uncompactified
dimensions, $v_2$ and $v'_2$ is the the volume of the 2-tori
in the $67$ and in the $89$ directions respectively;
$I$ and $I'$ refer to the fixed points of $R_1$ and $R_2$
respectively.

The chain of reasoning that determines the
solution is then as follows.
To cancel the tadpoles of the 8-forms from the untwisted
sector (the terms proportional to $\frac{v_6 v'_2}{v_2}$
and $\frac{v_6 v'_2}{v_2}$), we need
$32$ branes of each kind with
$\g_{1, 7}$ and $\g_{1, 7'}$ equal to $\1$,
and $\g_{\O_1, 7}$ and $\g_{\O_2, 7'}$ both
symmetric,  which can be chosen to be $\1$ with
a unitary change of basis of Chan-Paton indices.
One can then use the argument presented in \PolcII\
which considers the amplitude in which a closed-string twisted state
turns into open string states. Conservation of $\Omega_1$ and
$\O_2$ requires that $\g_{R, 7}$ and $\g_{R, 7'}$
both be symmetric, which in turn implies that
$\g_{\O_2, 7}$ and $\g_{\O_1, 7'}$ must also be symmetric.
This can be consistent only if we choose vacuum states
in the $77'$ to have $R=-1$ so that all oscillator
states with $\Omega^2 = -1$ are projected out.
Cancellation of the tadpoles of 6-forms
from the twisted sector (the terms in \tadpole
proportional to $v_6$) then determines
that the branes are distributed in groups of eight
at the fixed planes, with $\g_{R, 7}=\1$
and $\g_{R, 7'} =-\1$. This determines the
solution completely.

The next simplest orientifold is when the $K3$ is given by
$Z_3$ orbifold of a hexagonal lattice. In this case,
$z_l \equiv z_l + 1\equiv e^{2\pi/3} z_l, l=1,2$. The element
$\a$ in \orbi\ is given by $\a : (z_1, z_2) \rightarrow
(e^{2\pi/3} z_1, e^{-2\pi/3} z_2)$ and $\b$ is the same
as in \orbi. We are thus interested in the projection
$\frac{1}{6}( 1+ \a +\a^2) (1+\O (-1)^{F_L} R_2 )$
Now,  because $\OF$, in this case interchanges the sectors
twisted by $\a$ with those twisted by $\a^2$,
one can easily see that this orientifold is T-dual to the $Z_3$
orientifold with the usual $\O$ projection discussed in
\refs{\GiJoI, \DaPaII}. This model  has $10$ tensor multiplets
and $11$ hypermultiplets, and $32$ 7-branes of one kind.
If they are all located at the fixed point
of $R_2$, that is also invariant under $\a$, then
the gauge group is $SO(16)\times U(8)$ with hypermultiplets
in $(1, 28)+ (16, 8)$.

To find a potential F-theory dual on a Voisin-Borcea
orbifold, we consider the configuration in which
there are eight 7-branes at each fixed point of
$R_2$ so that the tadpoles are canceled locally. One fixed
point of $R_2$ is invariant under $\a$, and the remaining
three form a triplet. The gauge group is
$SO(8)\times SO(8)$ with one adjoint hypermultiplet under
the first $SO(8)$ that comes from the fixed points that
form a triplet under $\a$. To identify the F-theory dual
we need to find an elliptic Calabi-Yau $X$
with the right Hodge numbers.
The Hodge number can be calculated by
compactifying further on a $T^2$ and
computing the Type-IIA spectrum as in \MoVaII.
We then have
\eqn\something{h^{11}(X) = r(V) + T + 2,\quad h^{21} = H^0 -1,}
where $r(V)$ is the rank of the gauge group, $T$ is the
number of tensor multiplets, and $H^0$ is the number
of hypermultiplets that are uncharged with respect to
the Cartan subalgebra of the gauge group.
Thus, the candidate Calabi-Yau should have $h^{11}=20$
and $h^{21}=14$. Happily, there is a unique Voisin-Borcea
with the above Hodge numbers which corresponds to
$(r, a, \d ) =(11, 9, 1)$ in the notation of
\refs{\MoVaII, \Borc}. Indeed, this model has
the same matter content as the orientifold configuration with
local tadpole cancellation.

\bigskip
\leftline{ \secfont Acknowledgements}
\bigskip

We would like to thank Sunil Mukhi and Joe Polchinski
for valuable correspondence and discussions.
This work was supported in part by the U. S. Department of Energy
under Grant No. DE-FG03-92-ER40701.
\vfill
\eject
\listrefs
\end